\documentclass[superscriptaddress,twocolumn,aps,pra,longbibliography]{revtex4-2}
\usepackage{graphicx,hyperref,color}
\usepackage{amsfonts,units}
\begin{document}
\title{Future of Quantum Computing}
\author{Scott Aaronson\href{https://orcid.org/0000-0003-1333-4045}{\includegraphics[scale=0.05]{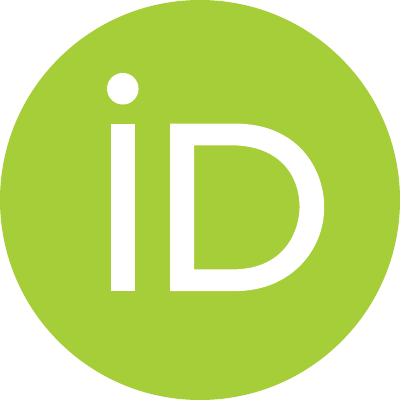}}}
\affiliation{Department of Computer Science, The University of Texas at Austin, TX 78712, USA}
\author{Andrew M. Childs\href{https://orcid.org/0000-0002-9903-837X}{\includegraphics[scale=0.05]{figures/orcidid.pdf}}}
\affiliation{Department of Computer Science, Institute for Advanced Computer Studies, and Joint Center for Quantum Information and Computer Science, University of Maryland, College Park, MD 20742, USA}
\author{Edward Farhi}
\affiliation{Google Quantum AI, Venice, CA 90291, USA}
\author{Aram W. Harrow\href{https://orcid.org/0000-0003-3220-7682}{\includegraphics[scale=0.05]{figures/orcidid.pdf}}}
\affiliation{Center for Theoretical Physics, Massachusetts Institute of Technology, Cambridge, MA 02139, USA}
\author{Barry C.\ Sanders\href{https://orcid.org/0000-0002-8326-8912}{\includegraphics[scale=0.05]{figures/orcidid.pdf}}}%
 \email{sandersb@ucalgary.ca}
\affiliation{Institute for Quantum Science and Technology, University of Calgary,
Alberta T3A~0E1, Canada}
\date{\today}
\begin{abstract}
On Tuesday 26\textsuperscript{th} November 2024,
four discussants participated in a moderated virtual panel titled `Future of Quantum Computing'
as one session
of the 8\textsuperscript{th} International Conference on Quantum Techniques in Machine Learning hosted by the University of Melbourne.
This article provides a detailed summary of the discussion in this lively session.
\end{abstract}
\maketitle
\section{Introduction}
\emph{Quantum Techniques in Machine Learning} is an annual international conference that focuses on the interdisciplinary field of quantum technology and machine learning, the objective of the conference being to gather leading academic researchers and industry players to interact through a series of scientific talks. The conference series focuses on the interplay between machine learning and quantum physics.
The 8\textsuperscript{th}
International Conference on Quantum Techniques in Machine Learning~\footnote{qtml2024.org}
was held at the University of Melbourne 25\textsuperscript{th}--29\textsuperscript{th}
November 2024.

As the conference dates overlapped with the United States Thanksgiving Holiday on Thursday 28\textsuperscript{th} 2024,
many USA-based scientists in this field indicated an inability to participate.
Turning a weakness of conference timing into a strength,
the conference programme and organising co-chairs decided to hold one virtual panel session comprising USA-based scientists as discussants to talk about the burning question of what is the future for quantum computing.

The panel was established and moderated by Barry Sanders.
The discussants were Scott Aaronson,
Andrew Childs,
Eddie Farhi, and
Aram Harrow.
The discussants introduced themselves and their perspectives on the future of quantum computing with an emphasis on machine learning,
as the overlap of machine learning and quantum computing is a key part of this conference.

This article is structured as follows.
Opening statements are presented in~\S\ref{sec:openingstatements}
with one subsection for each of the four discussants.
The ensuing dialogue is presented in~\S\ref{sec:dialogue}.
Conclusions are drawn in~\S\ref{sec:conclusions}
\section{Opening statements}
\label{sec:openingstatements}
Each discussant started with a brief personal introduction and remarks on the topic of the panel.
The order of speaking was the order of the discussants on the Zoom screen.
The opening statements presented here are slightly abbreviated and modified from what was said to improve clarity and readability.
\subsection{Scott Aaronson}
I am optimistic about quantum computing,
but I have been away from the field as I just took two years off to work on the OpenAI safety and alignment team, which no longer exists~\cite{Hay24}.
I have been asking `what have I missed'
over the past two years.
Although artificial intelligence (AI) is a giant that towers over everything else, 
if you took that away from the picture,
quantum computing is in an exciting time right now.

The year 2024 has just seen a genuine logical qubit that can outperform the underlying physical qubits and could be a building block of a future scalable system.
Two-qubit physics gates with 99.9\% fidelity in trapped ions and other systems have been achieved so we are close to or already at the threshold for fault tolerance, which was not true before.

Error rate as a function of year has gone down;
entirely realistic within the next decade doing quantum simulations that would be hard to reproduce with a classical computer and would at the very least be scientifically interesting to material science, chemistry, high-energy physics and other fields that care about quantum simulations.
Once the bar of `scientifically interesting' is cleared, then go to the higher bar of `commercially useful' such as battery or photovolta\"{i}c.
We do not have a clear winner between architectures such as trapped ion, neutral atoms, superconducting qubits, photonic qubits.
Very much still a live race.

As for algorithms, less dramatic progress but still very nice.
After 30 years, we still have not seen a breakthrough at the same level of Shor's factorisation algorithm~\cite{Sho94} and Grover's search algorithm~\cite{Gro97} in the 1990s.
Maybe that is an unfair standard as those algorithms could have been low-hanging fruits that have been picked.
The closest we have come to meeting that standard is work by fellow panelists;
kudos to them for that.
I am excited by Stephen Jordan and others adapting the Yamakawa-Zhandry breakthrough~\cite{YZ24} from a couple of years ago to get a new quantum algorithm that gets a better approximation ratio for some \textsc{NP-Hard} optimization problem with some algebra\"{i}c structure in it~\cite{JSW+24}.

There is more to be discovered with quantum algorithms.
The biggest applications are quantum simulation and breaking public-key encryption that currently protects the internet, and the latter is not necessarily a positive for the world.
The onus is on us to expand the list of applications.
Of course we know many Grover-type speed-ups~\cite{Gro97},
and it will take a very long time until those become a win in practice.
We know other heuristic algorithms, such as quantum-walk algorithms and QAOA~\cite{FGG14} and so forth that might sometimes give us speed-ups, but they remain subjects of active investigation.
\subsection{Aram Harrow}

One of the most exciting recent developments in computer science has been the dramatic progress of AI based on transformers~\cite{VSP+17}, diffusion models~\cite{GJvW24}, etc.  Unlike many past algorithms, these generally operate in regimes where they are not provably correct.  In some cases, there is not even an unambiguous description of what correctness would mean.

A simple example of this is gradient descent.  Although gradient descent can be proven to find a global minimum of a convex function~\cite{BV04},
it is often operated on non-convex functions where we do not expect it to find a global minimum.  Moreover, there is evidence that,
in machine-learning settings, the best solution from a generalization perspective comes from not exactly minimizing the empirical loss, but rather from running stochastic gradient descent for a limited amount of time.  This leads to a situation where theory lags far behind practice.  Ideas such as NP-completeness cannot answer the key questions.

This puts quantum machine learning in an uncomfortable position.
Studies of quantum algorithms has been a successful field, despite the lack of large-scale quantum computers on which to run these algorithms, because we can prove the correctness of algorithms.   Using proofs, we can be confident that Shor's algorithm~\cite{Sho94} works, that we can perform fault-tolerant quantum computing once gate errors are low enough~\cite{AB97},
and that we can build a universal simulator.
However,  proofs are not going to get us very far in understanding the promise of quantum machine learning.  There are some nontrivial results we can obtain about information-theoretic learnability, and about components of a larger algorithm, such as gradient descent.  But overall we have to grapple with the difficulty of developing heuristic algorithms without large machines on which to test them.

Another issue with quantum machine learning is the challenge of input data.  
If we had a quantum random access memory (qRAM)~\cite{GLM08}, then we could do a lot, but there are good arguments that this is unrealistic.  Scott Aaronson wrote a good critique of unrealistic input data models several years ago~\cite{aaronson2015},
and I described some strategies for how to pursue quantum speed-ups for large data sets without qRAM assumptions~\cite{Har20}.

Looking to the future, it would be interesting to see cases where it is useful to have a quadratic or heuristic speed-up.
Perhaps, they could be combined with exponential speed-ups in a way where the resulting end-to-end algorithm outperforms classical competitors.  If we combine quantum machine learning with tasks like quantum simulation, this could also be a realistic way to justify some of the powerful assumptions about quantum input that have sometimes been used.

\subsection{Andrew Childs}
I would like to emphasize that there is a lot of uncertainty about what quantum computers will be good at.
We have a couple of applications showing that quantum computers would offer an advantage for cryptanalysis and simulating quantum mechanics. Beyond that, there are hints and ideas to explore, but,
because we do not have large-scale quantum computers to try things out, we have to rely on proofs to understand the performance of quantum algorithms for the most part.
This definitely limits our understanding.

We know that,
for quantum computers to get significant speed-up, they need to take advantage of special structure.
They do not offer significant speed-up for unstructured problems,
so we do not expect to be able to just throw them at anything.
To get a significant advantage, there has to be something special about a problem that a quantum computer can latch onto.
We do not have a good understanding of what those structures look like.
We have some examples and some natural candidates to explore, but our understanding of where we expect to see quantum speed-up is definitely incomplete.

Quantum computing is exciting because there are many things to try.
We may eventually find quantum speed-ups for problems that we have not yet envisioned as good targets for quantum computers.
Hopefully, we will discover surprising applications as we get larger-scale devices.
Experimental advances have been very impressive in recent years, and it will be exciting to see how things change as we get larger devices to try things out on.
But
for now, it is unclear what the applications of quantum computers will be.

Quantum simulations are probably the most compelling application that we currently know of.
If we want to understand things that behave according to the rules of quantum mechanics,
this should be hard for classical devices and easy for quantum computers~\cite{Fey82}.
But even for quantum simulation, establishing a quantum advantage may not be straightforward because of sophisticated classical computational methods and theoretical tools for problems of practical significance in areas like chemistry and materials science.
Computational chemists can already do a lot with the tools they have at hand.

When we are trying to do something useful with a quantum computer,
we are up against the difficulty of building a coherent quantum device that can perform complicated calculations, and also the fact that classical algorithms can do very clever things.
This makes it complicated to understand where we can hope to see advantages from quantum computers. This area deserves a lot more exploration, but we should be cautious with our expectations about what we can accomplish.

\subsection{Edward Farhi}
My main interest is in the development of quantum algorithms.
I am eternally optimistic that a quantum computer might be good for optimisation.
In 2016, I started working at  Google as a visiting Research Scientist.  I am now a full-time employee. 
I am not a mouthpiece for Google,
but I am a Google employee.

The development of quantum algorithms is a tricky business because the usual standard is a performance guarantee.
Proving things in a computer science language often means showing things in worst-case,
but when I first went into quantum computing,
I started talking to Michael Sipser who said there is no such thing as the set of worst-case instances of anything.
I did not really know what that meant,
but, after a while, I kind of understood it.
If you identify what you think is a worst-case set,
you might be able to solve one of the so-called worst-case instances.
Thus, there is no such things as the set of worst-case instances,
yet that is the standard that is used by computer scientists in asserting that things are difficult as you want to show that something solves everything including the worst case.

I am intrigued by the possibility that quantum computers can solve problems in practice.
Peter Shor gives a talk about the development of classical algorithms,
and he talks about simulated annealing~\cite{vLA87}
as one of four examples of classical algorithms without provable guarantees that work extremely well.
It is very hard to prove that simulated annealing works because you have to show that there is some Markov chain whose gap is not exponentially small,
but it works in practice.

We should be open to the possibility that we can explore quantum algorithms without proving that they outperform the best known classical algorithms.  We can  still learn things by trying and in some cases we prove interesting things about the algorithms. We can also learn things by simulation.

Now the ultimate way we could learn things would be to run on a big computer, but, unfortunately, that is not going to happen very soon because,
even if you try to run things on an existing quantum computer,
the limited size and the noise mean that the performance even of a perfect algorithm will be degraded so it is hard to judge its performance. But ideally if we had an error-corrected device of sufficient size,
we could do experiments on it that go beyond what we can achieve by classical simulation or by classical analysis of a certain size. For example we can use tensor methods to predict performance.  Although this predicts how a large quantum computer could perform you still need the device to produce the associated strings.  I am generally open to the possibility of learning things about quantum algorithms without necessarily proving that they work in worst-case.

I think we have learned a lot of interesting things about quantum algorithm.
I am really a practitioner and I do not like to talk about whether I am optimistic or not.
I just decide what I am going to do,
and what I do is I work on things and that speaks to the fact that I think they're not crazy,
and that is good enough to keep me going.
I would like to say it is very hard to make predictions,
especially about the future,
so instead I can only talk about what I do, and what I do is attempt certain mathematical approaches to quantum algorithms which either offer performance guarantees or else demonstrate properties of the algorithms which are of interest to folks who care about algorithms.

I want to say something about simulation when we get fault-tolerant quantum computers on platforms that are not gate model based. 
Today Aram was talking about the Fermi-Hubbard model.  At MIT~\cite{PSS24},
 people have optical lattices with 600 wells in which they put fermions.
They can not put two with the same spin at the same site.
They measure properties of the Fermi-Hubbard model.
One can ask if what  they measure is simulable or not. But the technical achievement is huge.  They have solved the sign problem because they use actual fermions which obey the right statistics. This and other analog type simulators need to be further explored.

\section{Dialogue}
\label{sec:dialogue}
The dialogue is presented in this section.
Initially the moderator raised issues,
and the discussants responded.
Each issue discussed is denoted by a subsection,
and the discussants are separated by subsubsections here.
Occasionally one discussant interrupted (politely of course in every case) another discussant to make remarks.
These interruptions are presented as an indented paragraph with attribution to the discussant making that remark.
\paragraph{Moderator:}
Sanders raised the distinction between algorithms and heuristics, which has come up during the discussions in various ways.
A lot of people at this conference are working at the intersection between AI and quantum computing and are performing messy heuristic tests,
which can be expensive to gather sufficient data.
So in this messy world of heuristics and algorithms,
the question is at what point is a result convincing in showing quantum advantage.
\subsubsection*{Aaronson}
We have seen almost 20 years of people trying out heuristic algorithms orginally on D-Wave devices~\cite{CT14} and now on a wide variety of devices.
For even longer than that,
people have been running classical simulations to try to understand empirically the behaviour of quantum algorithms
beyond what we can prove.
I should say that that is an absolutely essential thing to do and we should do it.

The fundamental difficulty is that it is not necessary for heuristic algorithms to do well
Rather the best quantum heuristic has to beat the best classical heuristic.
What we want to see is not just that a quantum device is winning by some constant factor,
such as to find the ground state of this particular system.
It is no surprise that this system is good at finding its own ground state.
Want to see better scaling.

We want to see better performance
arising, for example, from quantum interference among entangled states in an exponentially large Hilbert space
and not because of any other reason.
It is hard to do science about this,
and our experience over a decade or two has underscored the difficulties.

Every day or week I am seeing papers on the arXiv about a quantum approach to such-and-such:
either quantum neural nets or quantum finance, 
and often it does not even reach the point of doing a comparison against the best classical algorithm that we know.
OK,
so that can be discounted right away as,
if the comparison is not done, then we are not asking the central question.
However, once the key question is asked, then what we want to see is better scaling behavior and ideally some story about what is going on.
This is a tunnelling effect like we saw in the quantum walk on the glued trees~\cite{CCD+03},
or this is a Grover-type speed-up~\cite{Gro97},
something that helps us to understand why, when we scale up to larger systems this speed-up will not go away but rather get better and better.
\subsubsection*{Farhi}
OK I am going to object to that.
I think it is just so narrow to have one criterion to decide if a paper is interesting being whether it beats the best classical.
I work on the QAOA~\cite{FGG14}, and really interesting things have been discovered. For example the QAOA has supremacy at its shallowest depth~\cite{FH19}.
It is an IQP circuit~\cite{LA24}.
Also if you look at the parameters,
which you use to optimise the QAOA,
they fall on universal curves~\cite{DGPG25}.
Now understanding why the parameters fall on universal curves, which we see in different settings, is itself interesting.  It is also useful for running the algorithm since you can pick the parameters without the classical outer loop search.

Observing universal curves does not address the issue of whether the output is better than classical
so there are tons of questions you can ask about algorithms which are not simply the tired old question, which I hear about,
which is does it or does it not beat the best classical algorithm.
There are other things of interest to people who pay attention to the field.
\subsubsection*{Aaronson}
Without doubt there are thousands of interesting questions that you could ask.
People are free to write papers about any of them.
When I am out there talking to people in the investment world or the funding world or journalists, every week I am dealing with people who see all these papers and have taken from them the impression that there must be a quantum advantage for all these tasks or else why are people writing all these things about it?
\begin{quote}
\emph{Farhi:}
That is your burden
that you feel you have to defend.
\end{quote}
That is indeed indeed my burden.
That is what I deal with.

I think we have to look at what is the sort of perception in the external world,
and quantum is a synonym for awesome.
Quantum computing is regarded as speeding up everything.
In particular,
we have quantum image recognition~\cite{LZX+20}
or quantum handwriting recognition with
quantum neural nets~\cite{ZGKS99}.
We have seen companies go public on claims about these kinds of application or pitched to investors on such claims.
Some companies say there are papers about it.

Thus, I think it is not enough,
when writing a paper,
to not say anything false.
Rather, in an environment like this,
one has to anticipate and head off the way that claims are going to be misinterpreted.
\begin{quote}
\emph{Farhi:}
I do not feel "Scott's Burden"  whatsoever.
I do not care what people take away from my papers. There are over 4000 citations to the QAOA.  I do not monitor them and I am happy that we have given folks something definite to work on.
I find it odd that you feel this is something you have to fight for and you are critical of other people's work which you have not read.
I just care that my papers are high quality and correct.  No one has ever challenged this.
\end{quote}
If some people do not feel the burden,
then that doubles the burden that other people have.

\subsubsection*{Harrow}
I want to come back a little to Barry's question.
If you want to argue that a heuristic quantum algorithm has really achieved a big speed-up, you need to be confident about the classical comparison.  This is easiest if this is a problem that people cared about and tried hard to solve before anyone tried to find quantum algorithms for it.  On the other hand, we want quantum algorithm researchers to innovate, which includes finding new problem areas where the speed-ups might be better. 

Another way we can believe in a practical speed-up is if our understanding is fortified by proofs involving related toy problems.  For example, we might prove things about tunneling or taking advantage of symmetry in models that are artificially simple, and then guess that the insights extend to more complicated realistic problems.

\subsubsection*{Childs}
From a pragmatic standpoint,
if a quantum computer solves a problem
that could not have been solved with a classical computer,
then the quantum computer has done something interesting.
If we are able to use a quantum computer to design some material that is useful---say, we make a high-temperature superconductor with some material that we designed using a quantum computer---it does not matter if we could have come up with an algorithm to understand that material with a classical computer;
it will have been a useful thing to do.

To get there, 
we need to understand what quantum computers are good at.
We are going to do that by looking at toy examples and coming up with explanations of why our quantum algorithms behave the way they do.
We should absolutely do that kind of work,
even though the eventual goal is to use quantum computers for something that people are really interested in doing:
to break some code that people tried unsuccessfully to crack with classical devices, 
or to solve some optimization problem that we were not able to answer classically.

I think it is unlikely that we will be able to do those things by fumbling around in the dark.
There need to be some guiding principles.
We need to have some understanding of
the structures that we can take advantage of.
Eventually, the way we will really know that quantum computers are useful is when
we use them to get an answer that we were not able to get another way.

\subsubsection*{Aaronson}
Hopefully all of us can agree that the ideal is
open exploration where everyone is
thinking about classical algorithms, both theoretically and empirically,
where relevant.
Everyone is also thinking as hard as they can about classical algorithms,
and they are trying to do the comparison in as scientific way as possible.

When there are strong financial incentives,
sometimes that tilts the scales
in a particular direction.
In science,
it is never enough to only say things that are true.
We have a higher standard:
you should tell the whole truth.
Sometimes when I am reading quantum papers that are boasting about how great something is,
I am asking not just
are they telling me the truth but are they telling me the whole truth.
What are the caveats?
Do I have to dig into the appendices to see what the caveats are?
I think it is on us to be open about it,
right there in the abstract
that the quantum algorithm only works on these structured instances,
or say explicitly that there is also a classical algorithm to do such-and-such.

It is true that sometimes you find a quantum algorithm more easily than you can find a classical algorithm that does the same thing even if they both exist.
You can say we have an additional resource besides
running time;
we have human brain time to come up with the algorithm.
However, I think we have to guard against 
what I have called the `stone soup'~\cite{Bro47} phenomenon~\cite{Aar13},
which is where everyone gets very excited about doing something with a quantum computer and,
for that reason,
they put massive effort into understanding some particular problem and therefore they make progress on the problem.
On the other hand,
if they had put equal effort into classical algorithms,
they also would have made progress.

We have to try to separate out the component that is really coming from the thing that got us all excited about quantum computing in the first place,
which is the ability to choreograph these patterns of interference
to get a speed-up over what could have been done with a classical computer.
The scientific challenge remains of finding the set of problems where that really helps the most.

Eddie said earlier that there is no such thing as a worst-case set of instances,
but you have to be careful there because there is such a thing as a worst-case distribution over instances 
or distribution over instances\dots
\begin{quote}
\emph{Farhi:} But that is going to get technical.
Well, I'm not so sure about that either.
\end{quote}
For computing the parity of~$n$ bits,
if the bits are random,
then the algorithm's performance on that
random ensemble is the same as what it is in the worst case.
And you know one can give an argument for that:
worst-case, average-case, \dots
\begin{quote}
\emph{Farhi:}
I guess I do not follow that.
Maybe it does not matter, but
I wrote the paper showing a quantum computer could not speed up the determination of parity. This is in the 
 oracle setting~\cite{FGGS98}.
\end{quote}
Yeah, I know, I know, that is why I mentioned it, yeah, I know, I know.
Figuring for which ensembles of instances and which distributions of instances you get a speed-up and, if so by how much and if so what principles of quantum algorithms are being exploited there.
We should see ourselves as working together on this scientific quest to try to understand 
all of those things.
\subsubsection*{Farhi}
What I do not resonate with is this constant use of words like ``understanding" and ``insight";
I do not understand the need to understand.
For example, we have results on how the QAOA performs on the
Sherrington-Kirkpatrick Model~\cite{FGGZ22}.
We analysed it on typical instances, and we prove theorems about the performance of QAOA at infinite size.
This in itself is an achievement. Furthermore, we
beat the best assumption-free classical algorithm at depth 11~\cite{FGGZ22}.

I have no idea how it works, zero.
I do not know; I have no idea.
Neither do any of my collaborators:
no insight, nothing, and yet I know it works.
I did the calculation with my collaborators and we see results.
\begin{quote}
\emph{Childs:} Wouldn't it be great to know how it works and somehow use that insight to do something else?
\end{quote}
Yeah, but I think understanding is an overrated concept.
I am not interested in understanding;
I am interested in calculation.
\subsubsection*{Aaronson}
I think that,
if we had a situation in quantum computing that is analogous to what now exists in generative AI---where we just have this utterly civilisation-changing suite of algorithms,
even thought no-one really understands how they work---then we would have the luxury to say just
`here it is; take it or leave it'.

However, I would say that this is not the situation that we are in with respect to quantum algorithms.
The situation in quantum algorithms is much more that we have speed-ups for very specialised problems.
Hence, we do not have the
luxury of disdaining understanding.
\begin{quote}
\emph{Farhi:} OK

\emph{Harrow:} Obviously we would like understanding, but it may never come.

\emph{Farhi:} I think understanding is a false idol;
I really do.
\end{quote}

\section{Responses to audience questions}
\subsubsection*{Maria Schuld}
One argument presented here has been that computational complexity arguments---that is,
the culture of quantum computing---are not very useful in a field like machine learning,
and I would almost argue not for many quantum applications in real life,
and I am absolutely surprised that you all agree on that, more or less and very happy to hear that.
Now you are bickering about the idea of a second argument 
that, since we can simulate things and have toy problems,
can we in heuristics prove speed-ups or that one algorithm performs better than the other best-known classical and so on and so forth.

Is not there a third argument that goes to understanding why something works, not why something is better.
To give an example, in machine learning,
we could show that a reasonably unique quantum routine can reveal features that are just interesting for generalization, and we can show this is interesting.
That would be an absolutely powerful argument that I do not see in any papers at the moment and that I think many more people should talk about and think about.
My last statement:
there is a lot of theory to machine learning that tells you what is interesting or not but it does not live in deep learning,
but it lives in statistics or in other parts of machine learning.
\paragraph{Aaronson:}
You said something about the culture of computational complexity not being useful.
I think some parts are and some parts are not.
The parts when you are trying to clarify what resources I am using and am I getting a benefit by using quantum compared to not using it,
which are questions
Bernstein and Vazirani~\cite{BV97}
and Simon~\cite{Sim97} and Shor~\cite{Sho94} asked 30 years ago,
Those remain very good questions to ask.
Of course you cannot answer everything you want on the basis of theorems and so at some point you also do numerical simulations.
You take any form of evidence that you can.
I think clearly defining what you mean by advantage is key such as
trying to quantify if this just about a constant factor or whether
a better scaling behaviour can be found.
I think these questions remain as relevant as they ever were.

\paragraph{Harrow:}
I think there is a role for theorems in machine learning.
We do not expect those theorems to explain fully what is going on in the algorithm even for something provable like a kernel method or regression.
You say the algorithm provably does solve this problem,
but then how the algorithm is going to do 
for some prediction
depends on the details of the data.

When you put the whole package together 
you are going to be running it 
outside the provable regime 
in the interesting cases even if it is not deep learning.
We rarely are given 
a bunch of points that are a linear model
plus Gaussian noise,
such that we know this is actually the correct description of the model.
For sure it is still worth looking at theorems,
and
maybe the question is then what is the role of the theorems.

I view theorems as having value by providing a solid understanding of a toy model.
This understanding enables educated guesses about an extrapolation to the situation in front of you,
so I think the toy model has its use even if the toy model does not literally describe what is in front of you because hopefully it presents some of the stylized features.
Thus, I think we are describing different parts of the same elephant here;
I do not think there is a dramatic difference.
\subsubsection*{Alex Nietner}
I want to connect to something that Andrew Childs said in his introduction.
There were two statements that are extremely important in the context of quantum computation in general that, on a superficial level, somehow seem in contradiction with one another.
I would like you to elaborate on those two statements.
The first statement is that structure is needed for quantum algorithms to be good,
and the second statement is that there is another regime including quantum chemistry where quantum computing is good because it is quantum.
This is sort of a generic area with some quantum problem with the generic statement that quantum is good.

In a superficial sense where these things seem somewhat contradictory, there is something generic,
but, on the other side,
generic is not really good for quantum.
Is the notion of genericness sort of wrong in a sense for this comparison,
or is that eventually quantum simulation is maybe average-case easy?
\paragraph{Childs:}
I would say that the class of quantum dynamics that is natural and that you can efficiently simulate with a quantum computer is structured and is special.
It is not like any dynamics of any arbitrarily connected quantum system can be simulated efficiently.
If you have a $k$-local Hamiltonian~\cite{KKR06},
% AMC: this reference is not so relevant to the point I was making
% ~\cite{AGIK09},
or a system of fermions interacting in a geometrically local way,
or something that actually describes a physically plausible system,
those are kinds of systems that you can simulate efficiently, and they do evolve in some structured way so I think there is structure present.
There is not just one kind of structure that quantum algorithms can take advantage of.
There are algorithms that make use of Fourier transforms, which are perhaps doing something quite different than what happens in the dynamics of
$k$-local Hamiltonians.
Fundamentally,
there is structure in both cases.

\paragraph{Aaronson:}
The deepest resolution of the tension you identified is that we expect quantum computers to give big speed-ups for relatively generic quantum problems,
for problems of simulating all sorts of $k$-local Hamiltonians for example, but the very fact that we are talking about a quantum problem has imported the whole structure of quantum mechanics.
The much harder thing, the thing that requires much more specialised problems, is when you want quantum speed-ups for purely classical problems and you want quantum speed-ups that have nothing to do with quantum mechanics.

If you are happy with a Grover-type speed-up~\cite{Gro97},
then the problem can be unstructured,
but, if you want an exponential speed-up,
then our experience so far has been that you want something like periodicity or some sort of hidden Abelian group structure or a very special graph like the glued-trees graph~\cite{CCD+03} or something else that is
specially designed to allow
destructive interference to happen on all the outcomes that you do not want and constructive interference only on the outcome that you do what.
This need for huge structure is very closely correlated with trying to solve a classical problem.

\paragraph{Harrow:}
The need for structure might reflect the need to be able to prove things.

\paragraph{Aaronson:}
Those are back to the research questions.
There are many quantum simulation problems in chemistry for which tractable classical simulations are not known.
They are a good deal less structured than the ones that show exponential speed-up
to the extent we can ever prove exponential speed-ups.

Within the black-box model,
or the query-complexity model,
we can actually prove some of the stuff.
We can look at unstructured problems
such as parity or the OR function or 
total Boolean functions~\cite{BBC01},
and we can say that there is no asymptotic quantum speed-up at all or the quantum speed-up that you get is only polynomial
whereas we can study problems involving structured oracles
like Simon's Problem~\cite{Sim97} or Shor's period-finding problem~\cite{Sho94}
or the glued-trees problem~\cite{CCD+03} and we can say the speed-up is exponential.

Andris Ambainis and I, a decade ago,
asked can you ever get an exponential speed-up while querying a random oracle for a decision problem~\cite{AA14}.
That question remains open;
it might eventually get settled.
The big breakthrough two years ago by Yamakawa and Zhandry~\cite{YZ24} came when,
without answering our question,
they changed
the question.
They said 
if we change from our focus from
a decision problem to an NP search problem,
then there is an exponential speed-up relative to a random oracle and they showed exactly how.
It was precisely in probing this boundary about how little structure can you have
in a black-box problem and still get an exponential speed-up.

\paragraph{Farhi:}
I think that is right,
but also I think another approach you can take is look at physics
and see whether there are phenomenona you can translate perhaps into some algorithmic advantage. 
For example,
quantum systems tunnel through wells.
You might ask yourself,
\begin{quote}
    Can I set up an optimization problem where there is a  barrier, and 
    quantum mechanically tunnel through it
    but classically cannot overcome the barrier?
\end{quote}
There are examples of that direction of research.

There are also the things Andrew Childs and I used to work on
regarding propagating through all kinds of structures.
You see that quantum things move faster than you
might have imagined.
Just take the hypercube,
and if you start in one corner of the hypercube,
and you quantum walk,
you end up on the other side in time~$\nicefrac\pi2$ 
whereas classically you would just get lost in the middle.
Can you turn that into some kind of advantage algorithmically? 
That is the kind of stuff Andrew and I worked on.
So I think that is another approach, 
which is to just look at the physics phenomena and ask if there something weird in physics that you can take advantage of in an algorithmic fashion.
\paragraph{Aaronson:}
Yeah, and your group has had more success on that than anyone else has.
\paragraph{Farhi:} Thank you.

\subsubsection*{Abolfazl Bayat}
My question is about heuristic quantum AI.
What is a fair comparison in that domain between quantum algorithms and classical algorithms?
We have small quantum computers and, if you think about deep learning, we have classical algorithms with millions of neurons.
If we just look at the accuracy, what is a fair comparison?
When can we say our quantum algorithms are decent?
\paragraph{Harrow:}
Part of the value proposition for a quantum computer has to be beating an unfair comparison.
Qubits are always going to be more expensive than classical bits.
That is why we always seek some kind of speed-up.
ideally, a small number of qubits could beat a much large number of classical bits.
Failing that, maybe you are happy with a quadratic or cubic or 
$n^{1.5}$~\cite{LMM+23}
or whatever improvement---some lesser than exponential speed-up---because, for various reasons,
you might be happy with that:
intellectual curiosity or maybe this is part of a larger pipeline with exponential speed-ups elsewhere in the pipeline.
For whatever reasons,
maybe this becomes a
valid part of the comparison,
but I would push back on whether fairness is something we should aim for in these comparisons.
\paragraph{Aaronson:}
We should not think about this at if some rules are laid down from on-high about what counts or does not count as demonstrating a speed-up for quantum machine learning.
Any type of evidence is fair game.
However, what I want to see is that some process is followed that could have led to the conclusion that there is 
no quantum speed-up for this task or with this class of quantum machine learning models.

I want to see that it is not that someone just wrote in the bottom line that we need to see that quantum is better or our paper cannot be published
so let us now go backwards from that to fill in the justification that is consistent with that.
I want to see a process where we actually tried as hard as we could to equalise what could be equalised,
study for example a
small quantum neural network
vs a small classical neural network
and is the quantum neural network getting an advantage that
could not, as Aram said, be easily replicated by just adding a few more neurons to the classical neural network.

Better yet,
if we could get some understanding of what is going on there,
why is that the case?
I think that is a very modest ask.
What is striking is the number of papers I have seen that do not even clear that very
low bar of just noticing and asking these questions.
\section{Conclusions}
\label{sec:conclusions}
This lively panel and the excellent questions at the end of the panel discussion provided a hard look at where quantum computing is at today and thoughts about the next stages.
Diverse views exist regaarding how to advance the field of research on quantum algorithms and applications to areas such as optimisation and maachine learning.

The most exciting parts of the discussion touched on the value of quantum algorithms without proven speed-ups and honesty in papers concerning this issue.
In Aaronson's words~\cite{Aar24}:
\begin{quote}
Part of the panel devolves into a long debate between me and Eddie about how interesting quantum algorithms are if they don’t achieve speed-ups over classical algorithms, and whether some quantum algorithms papers mislead people by not clearly addressing the speed-up question (you get one guess as to which side I took).
\end{quote}
As for the main question, `what is the future of quantum computing?',
the healthy debate and, at times, discord,
illustrate the deep questions facing the quantum computing community and the excitement in dealing with profound issues during the march towards scalable quantum computing.
\bibliography{qcomp}
\end{document}